\documentstyle{article}
\newcommand{\calC}{{\cal C}}
\newcommand{\calH}{{\cal H}}

\newcommand{\bE}{{\bf E}}
\newcommand{\bP}{{\bf P}}

\newcommand{\tL}{{\tilde L}}

\newcommand{\tr}{{\tilde r}}
\newcommand{\tx}{{\tilde x}}

\newcommand{\trace}{{\rm Tr}}

\newtheorem{theorem}{Theorem}[section]
\newtheorem{prop}[theorem]{Proposition}
\newtheorem{lemma}[theorem]{Lemma}
\newtheorem{corollary}[theorem]{Corollary}
\newtheorem{remark}[theorem]{Remark}

\newcommand{\proof}{\noindent {\em Proof:~}}  

\newcommand{\nn}{\nonumber} 
\newcommand{\real}{{\bf R}}

\def\bd{\begin{displaymath}}
\def\ed{\end{displaymath}}

\def\eqref#1{(\ref{#1})} 
\def\qed{\hbox{\hskip 6pt\vrule width6pt height7pt depth1pt
    \hskip1pt}\bigskip}  
\def\to{\rightarrow}

\def\runinend{\enspace}
\def\ackname{Acknowledgement\runinend}%
\def\acknowledgements{\par\addvspace{17pt}\rmfamily
\def\ackname{Acknowledgements\runinend}%
\trivlist\if!\ackname!\item[]\else
\item[\hskip\labelsep
{\bf\ackname}]\fi}%

\begin{document} 
\bibliographystyle{plain} 

\hfill{} 
 
\thispagestyle{empty}

\begin{center}{\bf \Large Fluctuations of the Entropy Production
in Anharmonic Chains} 
\vspace{5mm}

Luc  Rey-Bellet\footnote{Email: lr7q@virginia.edu.}, 
Lawrence E. Thomas\footnote{Email: let@virginia.edu. Supported in part by 
NSF Grant 980139}\\    
\vspace{5mm}

\begin{center}
{\small\it Department of Mathematics, University of Virginia \\ 
Kerchof Hall, Charlottesville VA 22903, USA \\ }
\end{center}

\end{center}

\setcounter{page}{1} 
\begin{abstract} We prove the Gallavotti-Cohen fluctuation theorem for a
model of heat conduction through a chain of anharmonic oscillators
coupled to two Hamiltonian reservoirs at different temperatures.

\end{abstract} 

\section{Introduction}\label{intro} 

The Gallavotti-Cohen fluctuation theorem refers to a symmetry in the
fluctuations of the entropy production in nonequilibrium statistical
mechanics. It was first discovered in numerical experiments of Evans,
Cohen and Morris \cite{ECM} and then discussed in
\cite{GC} in the context of {\em thermostated systems}. As a mathematical
theorem it was proved for Anosov dynamical systems \cite{GC,Ge}. Soon
thereafter the fluctuation theorem was discussed in the context of
{\em stochastic dynamical systems} first by Kurchan \cite{Ku} and
then, more systematically by Lebowitz and Spohn, and Maes
\cite{LeSp,Ma1}. In particular, Maes discovered a general formulation
of the fluctuation theorem in the context of space-time Gibbs measures
which covers both Markovian stochastic dynamics and chaotic
deterministic dynamics (via a Markov partition).  As a mathematical
theorem the fluctuation theorem is proven for quite general stochastic
models with {\em finite} state space, such as lattices gases in a
finite box. Relations for the free energy related to the 
fluctuation theorem have been also discussed in \cite{Ja,Cr}. 

Among the consequences of the fluctuation theorem is the {\em non-negativity}
of entropy production although the proof of its {\em positivity} is more
difficult and is so far proved only in particular examples \cite{EPR2,Ma3}. 
We also note that in the related context of open
systems, classical and quantum, the production of entropy is discussed
at a general level in \cite{Ru1,JP1,P}. Again the non-negativity of
entropy production is relatively easy to establish, while the strict
positivity has been established only in particular models \cite{EPR2,JP2}.

In this paper we consider an {\em open system} consisting of a finite
(but of arbitrary size) chain of anharmonic oscillators coupled at its ends
only to reservoirs of free phonons at positive and different
temperatures \cite{EPR1,EPR2,EH,RT1,RT2}. 
In particular our model is completely Hamiltonian and
its phase space is not compact.
 
In order to establish the fluctuation theorem, two ingredients are
needed: one needs to prove a {\em large deviation theorem} for the
ergodic average of the entropy production and establish a {\em
symmetry} of the large deviation functional. The second part is
usually relatively straightforward to establish, at a formal level, 
since it follows from a symmetry of the generator of the dynamics. 
This formal derivation for models related to ours can be found in 
\cite{LeSp} and \cite{Ma2}. 

The first part, proving the existence of the large deviation
functional, involves technical difficulties, in particular if the
phase space of the model is {\em not compact}. In this case large
deviation theorems are established provided the system satisfies very
strong ergodic properties (such as hypercontractivity) see
e.g. \cite{DS,DZ,Wu}. In addition the entropy production is in general
an {\em unbounded} observable while standard results of large
deviations apply only to bounded observables.

In this paper we show how to treat these difficulties in the model at
hand. The techniques we use are based on the construction of Liapunov
functions for certain Feynman-Kac semigroups and Perron-Frobenius-like
theorem in Banach spaces. We heavily rely on the strong ergodic
properties of our model established in \cite{EPR1,EPR2,EH} and
especially in \cite{RT2}.

The Hamiltonian of the model, as in \cite{EPR1}, has the form
\begin{equation}
H\,=\, H_B + H_S + H_I \,.
\label{h1}
\end{equation}
The two reservoirs of free phonons are described by wave equations in
$\real^d$ with Hamiltonian
\begin{eqnarray}
H_B\,&=&\, H(\varphi_L,\pi_L) +  H(\varphi_R,\pi_R) \,, \nn \\
H(\varphi,\pi)\,&=&\, \frac{1}{2} 
\int dx \,(|\nabla\varphi(x)|^2 + |\pi(x)|^2)\,, \nn
\end{eqnarray}
where $L$ and $R$ stand for the ``left'' and ``right'' reservoirs, 
respectively.
The Hamiltonian describing the chain of length $n$ is given by 
\begin{eqnarray}
H_S(p,q)\,=\, \sum_{i=1}^n \frac{p_i^2}{2} + V(q_1, \cdots, q_n)\,, \nn \\
V(q)\,=\, \sum_{i=1}^n U^{(1)}(q_i) + \sum_{i=1}^{n-1}
U^{(2)}(q_i-q_{i+1}) \,, \nn
\end{eqnarray}
where $(p_i,q_i) \in \real^d \times \real^d$ are the coordinates and
momenta of the $i^{th}$ particle of the chain. The phase space of the
chain is $\real^{2dn}$.  The interaction between the chain and the
reservoirs occurs at the boundaries only and is of dipole-type
\bd
H_I \,=\, q_1 \cdot \int dx\, \nabla \varphi_L(x)\rho_L(x)  + 
q_n \cdot \int dx \,\nabla \varphi_R(x)\rho_R(x)\,, 
\ed
where $\rho_L$ and $\rho_R$ are coupling functions (``charge
densities'').

Our assumptions on the anharmonic lattice described by $H_S(p,q)$ are
the following:
\begin{itemize}
\item {\bf H1 Growth at infinity}: The potentials $U^{(1)}(x)$ and
$U^{(2)}(x)$ are $\calC^\infty$ and grow at infinity like
$\|x\|^{k_1}$ and $\|x\|^{k_2}$: There exist constants $A_i$, $B_i$, and $C_i$,
$i=1,2$ such that
\begin{eqnarray}
\lim_{\lambda \rightarrow \infty}  \lambda^{-k_i}    U^{(i)}( \lambda x ) 
\,&=&\, A_i \|x\|^{k_i} 
\,, \nn \\
\lim_{\lambda \rightarrow \infty}  \lambda^{-k_i +1} \nabla U^{(i)}( 
\lambda x ) 
\,&=&\, A_i k_i 
\|x\|^{k_i-2}  x \,, \nn \\
\| \partial^2 U^{(i)}(x) \| \, &\le& \, 
( B_i + C_i V(x))^{1-\frac{2}{k_i}}  \,. \nn 
\end{eqnarray}
Moreover we will assume that
\bd
k_2 \,\ge \, k_1 \, \ge \, 2\,,
\ed
so that, for large $\|x\|$ the interaction potential $U^{(2)}$ is
"stiffer" than the one-body potential $U^{(1)}$. 

\item{\bf H2 Non-degeneracy}: The coupling potential between nearest
neighbors $U^{(2)}$ is non-degenerate: For
$x\in {\bf R}^d$ and $m=1,2, \cdots$, let $A^{(m)}(x): {\bf R}^d
\rightarrow {\bf R}^{d^{m}}$ denote the linear maps given by
\bd
(A^{(m)}(x) v)_{l_1 l_2 \cdots l_{m}} \,=\, 
\sum_{l=1}^d \frac{\partial^{m+1}U^{(2)}}{\partial x^{(l_1)} 
  \cdots \partial x^{(l_m)} \partial x^{(l)}}(x) v_l \,.
\ed
We assume that for each $x \in {\bf R}^d$ there exists $m_0$ such that 
\bd
{\rm Rank} ( A^{(1)}(x), \cdots A^{(m_0)}(x)) = d \,.
\ed
\end{itemize}

\begin{itemize} 
\item{\bf H3 Rationality of the coupling}: Let ${\hat \rho}_i$ denote
the Fourier transform of $\rho_i$.  We assume that
\bd
|{\hat \rho}_i(k)|^2 \,=\, \frac{1}{Q_i(k^2)}\,,
\ed
where $Q_i$, $i\in \{L,R\}$ are polynomials with real coefficients and 
no roots on the real axis.  
\end{itemize}

We introduce now the temperatures of the reservoirs by choosing
initial conditions for the reservoirs. The Hamiltonian of a reservoir
is quadratic in $\Psi\equiv(\phi,\pi)$, $H=\langle \Psi, \Psi
\rangle/2$, and therefore the Gibbs measure at temperature $T$, 
$d\mu_T(\Psi)$ is the Gaussian measure with covariance $T \langle
\cdot \,,\, \cdot \rangle$. To construct nonequilibrium steady states we
assume that
\begin{itemize}
\item The initial conditions $\Psi_L=(\phi_L,\pi_L)$ and 
$\Psi_R=(\phi_R,\pi_R)$ of the reservoirs are distributed according
the gaussian Gibbs measures $d\mu_{T_L}$ and $d\mu_{T_R}$ respectively.
\end{itemize}

In order to define the heat flow through the bulk of the crystal 
we consider the energy of the $i^{th}$ oscillator which we take to be
\begin{equation}
H_i =  \frac{p_i^2}{2} + U^{(1)}(q_i) + \frac{1}{2} 
\left(U^{(2)}(q_{i-1}-q_i) + U^{(2)}(q_{i}-q_{i+1})\right) \,.
\label{hi}
\end{equation}
Differentiating $H_i$ with respect to time, one finds that
\bd
\frac{dH_i}{dt} \,=\, \Phi_{i-1} - \Phi_{i}\,,
\ed
where 
\begin{equation}
\Phi_i\,=\,\frac{(p_{i} + p_{i+1})}{2} \nabla U^{(2)}(q_{i}-q_{i+1})
\label{fii}
\end{equation}
is the heat flow from the $i^{th}$ to the $(i+1)^{th}$ particle. 
We define a corresponding entropy production by 
\bd
\sigma_i \,=\, \left(\frac{1}{T_R} - \frac{1}{T_L}\right) \Phi_i \,,
\ed
where $T_R$ and $T_L$ are the temperatures of the reservoirs.

There are other possible definitions of heat flows and corresponding
entropy production that one might want to consider. One might, for
example, consider the flows $\Phi_L$, $\Phi_R$ at the boundary of the
chains, and define $\sigma_b=-\Phi_L/T_L - \Phi_R/T_R$, or one might
take other quantities as local energies. But using conservation laws 
it is easy to see that all these heat flows have the
same average in the steady state. Moreover we will show that all the
entropy productions have the same large deviations functionals: the
exponential part of their fluctuations are identical.

We denote $(p(t),q(t))=(p(t,p_0,q_0,\Psi_L, \Psi_R),q(t,p_0,q_0,\Psi_L,
\Psi_R))$ as the Ha\-miltonian flow generated by the Hamiltonian \eqref{h1}, 
and consider the ergodic average 
\bd
{\overline \sigma_i}^t \equiv \frac{1}{t}\int_0^t
\sigma_{i}(p(s),q(s))\, ds \,.
\ed
The quantity $\sigma_i(p(s),q(s))$ depends on both the initial
conditions of the chain  and  of the
reservoirs which, by assumption, are distributed according
to thermal equilibrium.  By the ergodic theorem proven in \cite{RT2} there 
exists a measure $d\nu$ on $\real^{2dn}$ such that 
\bd
\lim_{t \to \infty}  {\overline \sigma_i}^t \,=\, 
\int \sigma_i \,d\nu \,.
\ed
for all $(p_0,q_0)$ and $d\mu_{T_L}$ and $d\mu_{T_R}$ almost
surely. Moreover $\int \sigma_i \,d\nu \equiv \langle \sigma
\rangle_{\nu}$ is independent of $i$ and as shown in \cite{EPR2}
\bd
\langle \sigma \rangle_{\nu} \ge 0 {\rm ~and~}  
\langle \sigma \rangle_{\nu} = 0 {\rm ~if~and~only~if~} T_L=T_R \,.
\ed

Given a set $A \subset {\bf R}$, we say that the fluctuations of
$\sigma_i$ in $A$ satisfy the large deviation principle with large
deviation functional $I(w)$ provided
\begin{eqnarray}
&&\inf_{w \in {\rm Int(A)}} I(w)\,\le \, \liminf_{t \to \infty}- \frac{1}{t} 
\log \bP\{{\overline \sigma_i}^t \in A\} \,\le \,  \nn \\
&& \limsup_{t \to \infty} -\frac{1}{t} 
\log \bP\{{\overline \sigma_i}^t \in A\} \,\le \, \inf_{w \in {\rm Clos(A)}} 
I(w) \,. \nn 
\end{eqnarray}
The study of large deviations for $\sigma_i$ is based on the moment
generating functionals $e_i(\alpha)$ given by
\bd
e_i(\alpha) \,=\, \lim_{t \to \infty} - \frac{1}{t} \log 
\int d\mu_{T_L} d\mu_{T_R} e^{-\alpha \int_0^t \sigma_i(p(s),q(s))\, ds} \,.
\ed
The main technical result of this paper is 
\begin{theorem}\label{momgen} If 
\bd
\alpha \in \left( - \frac{T_{\min}}{T_{\max}- T_{\min}}\,,\, 1 + 
\frac{T_{\min}}{T_{\max}- T_{\min}} \right)\,,
\ed 
$e(\alpha) \equiv e_i(\alpha)$ is finite and independent of $i$ and 
the initial conditions $(p_0,q_0)$.  Moreover $e(\alpha)$ satisfies the
relation
\bd
e(\alpha) = e(1-\alpha)\,.
\ed
\end{theorem}

As an application of the G\"artner-Ellis Theorem, see \cite{DZ}, 
Theorem 2.3.6, we obtain the Gallavotti-Cohen fluctuation theorem.
\begin{theorem}\label{galco} There is a neighborhood $O$ of the interval 
$[-\langle \sigma \rangle_{\nu}\,, \langle \sigma \rangle_{\nu}]$ such
that for $A \subset O$ the fluctuations of $\sigma_i$ in $A$ satisfy
the large deviation principle with a large deviation functional $I(w)$
obeying
\bd
I(w)-I(-w)\,=\, -w\,,
\ed
i.e., the odd part of $I$ is linear with slope ${-1/2}$.  
\end{theorem}
Theorem \ref{galco} provides information on the ratio of the
probabilities of observing the entropy production to be $w$ and $-w$:
roughly speaking we have
\bd
\frac{ \bP \{ {\overline \sigma_i}^t \in (w-\epsilon , w +\epsilon) \} } 
{ \bP \{ {\overline \sigma_i}^t \in (-w -\epsilon , -w +\epsilon) \} } \, 
\sim e^{wt}\,.
\ed

\section{Fluctuations of the entropy production}

\subsection{Exponential mixing and compactness}

As shown in \cite{EPR1,RT2}, under condition {\bf H3} the dynamics of
the complete system can be reduced to a Markov process on the extended
phase space consisting of the phase space of the chain $\real^{2dn}$
and of a finite number of auxiliary variables which we denote as
$r$. In the simplest case which we consider here, (corresponding to 
${\hat \rho}(k)\sim (k^2+\gamma^2)^{-1}$), $r=(r_1,r_n) \in
\real^{2d}$ and the resulting equations of motion take the form
\begin{eqnarray}
{\dot q} \,&=&\, p \,,\nn \\
{\dot p} \,&=&\, -\nabla_q V - \Lambda^T r \,, \nn \\
{  dr} \,&=&\, (-\gamma r + \Lambda p)\,dt + (2\gamma T)^{1/2} d\omega \,. 
\label{e23}
\end{eqnarray}
Here $p=(p_1,\cdots, p_n)$ and $q=(q_1,\cdots, q_n)$ denote the
momenta and positions of the particle, $r=(r_1,r_n)$ are the auxiliary
variables and $\omega$ is a standard $2d$-dimensional Wiener process. 
The linear map $\Lambda : \real^{dn} \to \real^{2d} $ is
given by $\Lambda (p_1, \ldots, p_n)= (\lambda p_1, \lambda p_n)$ and
$T : \real^{2d} \to \real^{2d} $ by $T(x, y) = (T_1 x, T_n y)$. Here
$T_1 \equiv T_L$ and $T_n\equiv T_R$ are the temperatures of the 
reservoirs attached to the first and $n^{th}$ particles respectively, 
$\gamma$ is the constant appearing in ${\hat \rho}$ and $\lambda$ is a 
coupling constant equal to $\|\rho\|_{L^2}$. 

The solution of Eq. \eqref{e23}, 
$x(t)=(p(t),q(t),r(t))$ with $x\in X=\real^{2d(n+1)}$ 
is a Markov process. We denote $T^t$ as the corresponding
semigroup
\bd
T^tf(x) \,=\, \bE_x [ f(x(t)]\,,
\ed
with generator
\begin{equation}\label{generator}
L\,=\, \gamma \left( \nabla_r T \nabla_r - r \nabla_r\right) + 
\left( \Lambda p \nabla_r - r \Lambda \nabla_p \right) + 
\left( p \nabla_q - (\nabla_q V(q)) \nabla_p\right)\,,
\end{equation}
and we denote $P_t(x,dy)$ as the transition probability of the Markov 
process $x(t)$. 
In \cite{RT2} we proved that the Markov process $x(t)$ has smooth
transition probabilities, in particular it is {\em strong Feller}, and
that it is (small-time) {\em irreducible}: For any $t>0$, any $x\in X$
and any open set $A\subset X$ we have $P_t(x,A)>0$.

There is a natural energy function associated to Eq.\eqref{e23}, 
given by
\bd
G(p,q,r)\,=\, \frac{r^2}{2} + H(p,q)\,,
\ed
which we employ throughout our discussion. In \cite{RT2} we have
constructed a {\em Liapunov function} for $x(t)$ from $G$: Let $t>0$
and $0<
\theta < \max(T_1,T_n)^{-1}$. Then there exists $E_0$ such that for
all $E> E_0$ there exist functions $\kappa=\kappa(E) <1$ and $b=b(E)
<\infty$ such that
\begin{equation}\label{lb}
T^t e^{\theta G}(x) \,\le \, \kappa(E) e^{\theta G}(x) + b(E)
{\bf 1}_{\{G\le E\}}(x)\,.
\end{equation} 
Moreover $\kappa(E)$ can be made arbitrarily small by choosing $E$
sufficiently large, in fact there exist positive constants
$c_{1}=c_{1}(\theta,t)$ and $c_{2} = c_{2}(\theta,t)$ such that
\begin{equation}\label{kappae}
\kappa(E) \leq c_{1}e^{-c_{2}E^{2/k_2}}.  
\end{equation}

By results of \cite{MeTw} it is also shown in \cite{RT2} that the
convergence to the unique stationary state, denoted by $\mu$, occurs
exponentially fast: Let $\calH_{\infty,\theta}$ denote the Banach space
$\{f\,;\, \|f\|_{\infty,\theta}
\equiv \sup_x |f(x)| e^{-\theta G(x)} < \infty \}$. Then there exist constants 
$r>1$ and $R< \infty$
\begin{equation}\label{expconv}
|T^tf(x) - \int f d\mu| \le R r^{-t} \|f\|_{\infty,\theta} e^{\theta G(x)}\,,
\end{equation}
which means that $T^t$, acting on $\calH_{\infty,\theta}$ has a spectral gap.
The methods of \cite{MeTw} are probabilistic and rely on a nice
probabilistic construction called splitting as well as coupling
arguments and renewal theory.

Under the condition given here, by taking advantage of the fact that
the constant $\kappa$ in the Liapunov bound \eqref{lb} can be made
arbitrarily small (this is not assumed in \cite{MeTw}), we can prove
stronger ergodic properties and also give a direct analytical proof of 
Eq. \eqref{expconv}.  

Besides the Banach space $\calH_{\infty,\theta}$ defined above we also
consider the Banach space $\calH^0_{\infty,\theta} =\{ f, |f|e^{-\theta
G}\in C_0(X)\}$ with  norm $\|\cdot\|_{\infty,\theta}$ ( $C_0(X)$ denotes 
the set of continuous functions which vanish at
infinity). Furthermore for $1\le p < \infty$ we consider the family of
Banach spaces $\calH_{p,\theta}= L^p(X, e^{-p\theta G(x)}dx)$ and
denote $\|\cdot\|_{p,\theta}$ the corresponding norms.

\begin{theorem} \label{compact} If $0 < \theta T_i < 1$, 
the semigroup $T^t$ extends to a strongly continuous quasi-bounded
semigroup on $\calH_{p,\theta}$, for $1\le p < \infty$ and on
$\calH^0_{\infty,\theta}$. For any $t>0$, $T^t$ is compact on
$\calH_{p,\theta}$, for  $1< p\le \infty$ and on $\calH^0_{\infty,\theta}$.
\end{theorem}

As an immediate consequence of the spectral properties of positive semigroups 
\cite{Gr} and the irreducibility of $x(t)$ we have 

\begin{corollary} The Markov process $x(t)$ has a unique invariant measure 
$d\mu$ and Eq. \eqref{expconv} holds. 
\end{corollary}

\proof: Since $T^t$ is a Markovian, compact, and irreducible semigroup 
the eigenvalue $1$ is simple with the constant as the
eigenfunction. This shows that the Markov process $x(t)$ has a unique
invariant measure. Moreover by the cyclicity properties of the
spectrum of a positive semigroup \cite{Gr}, and by the compactness of
$T^t$, there are no other eigenvalues of modulus
$1$. Eq. \eqref{expconv} follows immediately. \qed

\noindent
{\em Proof of Theorem \ref{compact}} In \cite{RT2}, Lemma 3.6, we
showed that for some constant $C$ $T^te^{\theta G} \le e^{ct}e^{\theta
G}$ provided $\theta T_i <1$ (see also Lemma \ref{welldef}
below). Therefore for $f$ $C^\infty$ with compact support we have,
using Ito's and Girsanov's formulas
\begin{eqnarray}
e^{-\theta G}T^t e^{\theta G}f(x)\,&=&\,\bE_x \left[ e^{\theta (G(x(t))-G(x))}
f(x)\right] \nn \\   
\,&=&\, \bE_x\left[ e^{\theta\int_0^t \gamma( \trace(T)-r^2)\,ds +\theta
\int_0^t \sqrt{2\gamma T} r d\omega(s)} f(x(t))\right] \nn \\
\,&=&\, \bE_x\left[ e^{ \gamma \theta \trace(T) + \gamma \tr(\theta^2 T -
\theta)\tr}f(\tx(t))\right] \,, \nn 
\end{eqnarray}
where $\tx$ is the process with generator 
\bd
\tL_\theta\,=\, L + 2\gamma \theta rT\nabla_r\,.
\ed
A computation shows that $\tL^T_\theta 1= \gamma 
\trace(1-2\theta T)$. Standard arguments show then that the semigroup 
associated with the process $\tx$ extends to a quasi-bounded and
strongly continuous semigroup on $L^p(dx)$, $1\le p < \infty$ and on
$C_0(X)$. Using the assumption that $\theta T_i <1$ and Feynman-Kac
formula we see that $e^{-\theta G}T^t e^{\theta G}$ extends too to a
quasi-bounded and strongly continuous semigroup on $L^p(dx)$, $1\le p <
\infty$ and on $C_0(X)$. This implies immediately that $T^t$ extends to a 
strongly continuous semigroup on $\calH_{p,\theta}$, $1\le p < \infty$
and $\calH^0_{\infty,\theta}$. The computation above also shows that
$T^t$ extends to a quasi-bounded semigroup on $\calH_{\infty,\theta}$.

We first prove the compactness of $T^t$ for $\calH_{\infty,\theta}$. If $f \in
\calH_{\infty,\theta}$ then $|f(x)| \le
\|f\|_{\infty,\theta} e^{\theta G(x)}$ and by \eqref{lb} and \eqref{kappae} 
we obtain
\begin{eqnarray}
| {\bf 1}_{G\ge E} T^t f(x)| \,&\le& \, e^{\theta G(x)} 
\sup_{\{y:G (y)\geq E\}} \frac{|T^tf(y)|}{e^{\theta G(y)}} \nn \\
\,&\le& \, e^{\theta G(x)} \|f\|_{\infty,\theta} \sup_{\{y:G (y)\geq E\}} 
\frac{T^te^{(\theta G(y))}}{e^{\theta G(y)}} 
\nn \\
 \,&\le& \,\kappa(E) e^{\theta G(x)} \|f\|_{\infty,\theta}\,. \label{fg}    
\end{eqnarray} 
From the bounds \eqref{fg} and \eqref{kappae} we conclude that the
operator ${\bf 1}_{\{G\ge E\}} T^t$ converges uniformly to $0$ in
$\calH_{\infty,\theta}$ as $E\to \infty$. The semigroup $T^{t}$ has a
$C^{\infty}$ kernel since it is generated by a hypoelliptic operator
see \cite{RT2}, Proposition 4.1, so, by the Arzela-Ascoli theorem 
${\bf 1}_{\{G\leq E\}} T^{t/2}{\bf 1}_{\{G\leq E\}}$ is
compact, for any $E$.  Therefore we obtain
\bd
T^{t}= \lim_{E\rightarrow\infty}{\bf 1}_{\{G\leq E\}}T^{t/2}{\bf 1}_{\{G\leq
E\}}T^{t/2} \,,
\ed
where the limit is in the norm sense from \eqref{fg} above, i.e., 
$T^{t}$ is the uniform limit of compact operators, hence is compact. 
 
The compactness of $T^t$ for $\calH^0_{\infty,\theta}$ follows from
the same argument. In fact by Eq.\eqref{kappae}, for any $t>0$,  
$T^t \calH_{\infty,\theta} \subset \calH^0_{\infty,\theta}$.

To prove the compactness of $T^t$ on ${\mathcal H}_{p \theta}$, 
$1 <  p < \infty$, we note that 
\begin{eqnarray}
|T^{t}f (x)|&=& |{\bf E}_{x}[f (x (t))]| \nn \\
 &=& |{\bf E}_{x}[e^{\frac{\theta}{q} G (x(t))}e^{-\frac{\theta}{q} 
   G(x(t))} f (x (t))]|\nonumber\\ 
   &\leq& \left({\bf E}_{x}[e^{\theta G (x(t))}]\right)^{1/q} 
\left({\bf E}_{x}[e^{-\frac{p \theta}{q} G(x(t))}f^{p} (x(t))] \right)^{1/p}.
\nn
\end{eqnarray}
Thus using the bound \eqref{kappae} and the fact that $T^t$ is quasi-bounded 
on $\calH_{1,\theta}$ we obtain
\begin{eqnarray}
\|{\bf 1}_{G\geq E}T^{t}f\|_{\theta,p}^{p}  &\le& \!\!\!\!\!
\int_{\{x:G (x)\geq E\}}\!\!\!\!\!\!\! 
{\bf E}_{x}[e^{\theta G (x(t))}]^{\frac{p}{q}}\,\, 
{\bf E}_{x}[e^{-\frac{p \theta}{q}  G(x(t))}f^{p} (x(t))] 
e^{-p \theta G(x)} dx \nonumber \\
&\leq& \sup_{\{x: G (x)\geq E \}}\left( 
\frac{{\bf E}_{x}[e^{\theta G (x(t))}]}{ e^{\theta G(x)}}\right)^{\frac{p}{q}}
\|T^{t} (e^{-\frac{ p \theta}{q} G} f^{p})\|_{1,\theta} 
\nn \\
&\leq & \kappa(E)^{\frac{p}{q}} e^{ct}
\|e^{-\frac{p \theta}{p} G}f^p  \|_{1,\theta}  \nn \\
&=& \kappa(E)^{\frac{p}{q}}  e^{ct}   \|f\|_{{\theta,p}}^{p}\,. \nn
\end{eqnarray}

As in the case $p=\infty$, we conclude from the bound \eqref{kappae}
that the operator ${\bf 1}_{G\ge E} T^t$ converges uniformly to $0$ in
$\calH_{p,\theta}$ as $E\to \infty$. Using that the kernel of 
${\bf 1}_{\{G\le E\}} T^t {\bf 1}_{\{G \le E\}}$ is bounded, 
we conclude that $T^t$ is compact on $\calH_{p,\theta}$ for $1< p <\infty$. 
\qed

\subsection{Heat flow and generating functionals}

In order to define the heat flows we note that we have
\bd 
\frac{d}{dt} T^t H\,=\, LT^t H \,=\, T^t(  - r \Lambda p ) \,=\, T^t 
(-\lambda r_1p_1 - \lambda r_n p_n)\,.
\ed
Hence we identify $\Phi_0 \equiv -\lambda r_1p_1$ as the observable
describing the heat flow from the left reservoir into the chain and
$\Phi_n \equiv \lambda r_np_n$ as the heat flow from the chain into
the right reservoir. As in the introduction we define the energy $H_i$
of the $i^{th}$ oscillators by Eq.\eqref{hi}, for $i\le 2\le n-1$, and  
\begin{eqnarray}
H_1 \,&=&\, \frac{p_1^2}{2} + U^{(1)}(q_1) + \frac{1}{2} U^{(2)}(q_1-q_2) \,,
\nn \\
H_n \,&=&\, \frac{p_n^2}{2} + U^{(1)}(q_n) + \frac{1}{2} U^{(2)}(q_{n-1}-q_n)
\,. \nn
\end{eqnarray}
With the heat flows $\Phi_i$, $i=1,\cdots, n$, defined as in
Eq. \eqref{fii} we have
\bd
LH_i\,=\, \Phi_{i-1}-\Phi_i\,, \quad i=1,\cdots, n\,.
\ed
and we define the entropy productions $\sigma_i$, $i=0,\cdots,n$ by 
\bd
\sigma_{i}\,=\,  \left(\frac{1}{T_1} - \frac{1}{T_n}\right) \Phi_i\, 
\quad i=0, \cdots, n \,.
\ed

We now provide several identities involving the generator of the
dynamics and the entropy production, which will play
a crucial role in our subsequent analysis.

\begin{lemma} \label{simple} 
Let the function $R_i$, $i=0,\cdots, n$ be given by
\begin{equation}\label{ri}
R_{i}\,=\, \frac{1}{T_1} \left(\frac{r_1^2}{2} + \sum_{k=1}^{i}
H_i(p,q)\right) + \frac{1}{T_n} \left( \sum_{k=i+1}^{n} H_i(p,q) +
\frac{r_n^2}{2} \right) \,.
\end{equation}
Then we have
\begin{equation}
\sigma_{i} = r T^{-1} r - \trace(I) +  L R_{i} \,.\label{magic}
\end{equation}
\end{lemma}

\proof This is a straightforward computation. \qed

\begin{remark}{\rm 
This shows that, up to a derivative, all the entropy productions are
equal to the quantity $rT^{-1}r - \trace{\gamma I}$ which is
independent of $i$ and involves only the $r$-variables. 
}
\end{remark}

Let $L^T$ be the formal adjoint of the operator $L$ given by
Eq. \eqref{generator}
\begin{equation}\label{adgenerator}
L^T\,=\, \gamma \left( \nabla_r T \nabla_r + r \nabla_r\right) - 
\left( \Lambda p \nabla_r - r \Lambda \nabla_p \right) -
\left( p \nabla_q - (\nabla_q V(q)) \nabla_p\right)\,,
\end{equation}
and let $J$ be the time reversal operator which changes the sign of
the momenta of all particles, $Jf(p,q,r)=f(-p,q,r)$. 

The following identities can be regarded as  operator identities on
$\calC^\infty$ functions. That the left and right side of Eq. \eqref{conj}  
actually generate semigroups for some non trivial domain of $\alpha$ is a 
non trivial result which we will discuss in Section \ref{fksemi}.

\begin{lemma} \label{detbal} We have the operator identities
\begin{equation}\label{conj0}
e^{R_i} J L^T J e^{-R_i}\,=\, L - \sigma_i  \,,
\end{equation}
and also for any constant $\alpha$  
\begin{equation}\label{conj}
e^{-R_i} J ( L^T - \alpha \sigma_i) J e^{R_i}\,=\, L - (1-\alpha) \sigma_i  \,.
\end{equation}
\end{lemma}

\proof We write the generator $L$ as $L=L_0+L_1$ with 
\begin{eqnarray}
L_0\,&=&\, \gamma \left( \nabla_r T \nabla_r - r \nabla_r\right)  \label{l1} \\
L_1\,&=&\,  \left( \Lambda p \nabla_r - r \Lambda \nabla_p \right) + 
\left( p \nabla_q - (\nabla_q V(q)) \nabla_p\right) \label{l0}\,.
\end{eqnarray}
Since $L_1$ is a first order differential operator we have
\bd
e^{- R_i} L_1  e^{ R_i}\,=\, L_1 + L_1 R_i  \,=\, 
L_1 +  \sigma_i\,.
\ed
Using that $\nabla_r R_i = T^{-1}r$ we obtain  
\begin{eqnarray}\label{ww2}
e^{- R_i} L_0  e^{ R_i}
\,&=&\,e^{- R_i} \gamma (\nabla_r -T^{-1}r)T\nabla_r e^{ R_i} \nn \\
\,&=&\, \gamma \nabla_r T (\nabla_r + T^{-1}r) \,=\, L_0^T \nn \,.
\end{eqnarray} 
This gives
\bd
e^{- R_i} L  e^{ R_i}\,=\, L_0^T + L_1 + \sigma_i \,=\, J L^T J + \sigma_i\,,
\ed
which is Eq. \eqref{conj0}. Since $J\sigma_iJ=-\sigma_i$, 
Eq. \eqref{conj} follows immediately from Eq. \eqref{conj0}.  

\begin{remark}{\rm 
In the equilibrium situation, i.e., for $T_1=T_n=T$, Eq. \eqref{conj} is 
\bd 
e^{G/T} J L^T J e^{-G/T}\,=\, L \,,
\ed
which is simply {\em detailed balance}. Eq. \eqref{conj} can be interpreted 
in path space in the following manner \cite{Ma1}: 
Let $\Pi$ denote the time-reversal in path space on the time interval 
$[0,t]$: $\Pi(p(s),q(s),r(s))= (-p(t-s),q(t-s),r(t-s))$ and let 
$dP$ denote the measure on $\calC([0,t],X)$ induced by $x(t)$. Then Eq. 
\eqref{conj} implies that 
\bd
\frac{dP \circ \Pi}{dP} \,=\, e^{ R_i(x(t))-R_i(x(0))- \int_0^t 
\sigma_i(x(s))\,ds}\,.
\ed   
This formula exhibits the fact that the lack of microscopic
reversibility is intimately related to the entropy production. 
}
\end{remark}

We now turn to the study of the large deviations. As shown in 
\cite{RT2} the Markov process $x(t)$ is ergodic. In order to study the large 
deviations of $t^{-1} \int_0^t \sigma_i (x(s)) ds$ we consider the 
moment generating functionals 
\bd
\Gamma^i_x(t,\alpha) \,=\, \bE_x \left[ 
e^{- \alpha \int_0^t \sigma_i (x(s))\, ds } \right] \,.   
\ed 
Formally the Feynman-Kac formula gives
$\Gamma^i_x(t,\alpha)=e^{t(L-\alpha\sigma_i)}1(x)$, but since
$\sigma_i$ is not bounded, nor even relatively bounded by $L$, it is not 
obvious that $\Gamma^i_x(t,\alpha)$ exists for $\alpha \not=0$. 
Our goal is  to prove that 
$\Gamma^i_x(t,\alpha)$ exists and that the limit 
\begin{equation}\label{limit}
e(\alpha) \equiv \lim_{t \to \infty} -\frac{1}{t} \log \Gamma^i_x(t,\alpha)
\end{equation}
exists and  is finite in a neighborhood of the
interval $[0,1]$, and is independent of $i$ and of the initial
condition $x$.

The technical difficulty in proving the existence of the limit
\eqref{limit} lies in the fact that the functions $\sigma_i$ are {\em
unbounded}. Standard large deviation theorems for Markov processes (see
e.g. \cite{DS,DZ,Wu}) are proven usually under strong ergodic
properties for bounded functions and are not directly applicable.
Large deviations for unbounded functions are considered in
\cite{BM} for discrete time countable state space Markov chains under
conditions which amount in our case to $\sigma =o(G)$. In our case
this is clearly not satisfied since, in general $\sigma$ is {\em not}
bounded by $G$. 

But the $\sigma_i$ are very special observables, in particular
they are intimately linked with the dynamics as shown by the identities
Eqs.\eqref{conj0} and \eqref{conj}. The next lemma displays another identity 
which will be important in our analysis. 

\begin{lemma}\label{truc} We have the identity 
\begin{equation}\label{op}
L-\alpha \sigma_i \,=\, e^{\alpha R_i} {\overline L}_\alpha e^{-\alpha R_i}\,,
\end{equation}
where 
\begin{equation}\label{lba}
{\overline L}_\alpha \,=\, {\tilde L}_\alpha - 
\left((\alpha - \alpha^2)\gamma  rT^{-1} r - \alpha \trace(\gamma I)\right) 
\end{equation} 
and 
\begin{equation}\label{lta}
{\tilde L}_\alpha\,=\, L + 2\alpha \gamma r \nabla_r \,.
\end{equation}
\end{lemma}

\proof As in Lemma \ref{detbal} we write the generator $L$ as $L=L_0+L_1$, see 
Eqs.\eqref{l0} and \eqref{l1}.  
Since $L_1$ is a first order differential operator we have
\begin{equation}\label{w1}
e^{- \alpha R_i} L_1  e^{\alpha R_i}\,=\, L_1 +\alpha (L_1 R_i) \,=\, 
L_1 +\alpha \sigma_i\,.
\end{equation}
Using that $\nabla_r R_i = T^{-1}r$ is independent of $i$ we find that 
\begin{eqnarray}\label{w2}
e^{- \alpha R_i} L_0  e^{\alpha R_i}\,&=&\, 
\gamma\left( (\nabla_r + \alpha T^{-1}r)T (\nabla_r + \alpha T^{-1}r) - 
r(\nabla_r + \alpha T^{-1}r) \right) \nn \\
\,&=& \, L_0 +  \alpha\gamma ( r \nabla_r + \nabla_r r ) + 
(\alpha^2-\alpha)\gamma r T^{-1}r \nn \\
\,&=&\,  L_0 +  2 \alpha\gamma  r \nabla_r + (\alpha^2-\alpha)\gamma r T^{-1}r 
+ \alpha \trace{\gamma I}\,. 
\end{eqnarray}
Combining Eqs. \eqref{w1} and \eqref{w2} gives the desired result. \qed

\begin{remark}{\rm 
The identity \eqref{op} shows that all operators $L-\alpha \sigma_i$
are conjugate to the same operator ${\overline L}_\alpha$. This will
be the key element to prove that  $e(\alpha)$ is independent of
$i$. Furthermore it can be seen from Eqs. \eqref{lba} and \eqref{lta}
that ${\overline L}_\alpha$ has the form of $L$ plus a perturbation
which is a quadratic form in $r$ and $\nabla_r$. Such a perturbation
is indeed nicer than $\alpha \sigma_i$.  Also it should be noted that
${\tilde L}_\alpha$ has very much the same form as the operator $L$:
they differ only by the coefficient in front of the term
$r\nabla_r$. This fact will allow us to use several results on $L$ obtained
in \cite{RT2}.
}
\end{remark}

\subsection{Liapunov Function for Feynman-Kac Semigroups}\label{fksemi}

At this point we begin the study of ${\overline L}_\alpha$ as the
generator of a semigroup.

\begin{prop} \label{welldef} If $\theta$ and $\alpha$ 
satisfy the condition
\begin{equation}\label{condat}
-\alpha < \theta T_i < 1- \alpha \,,
\end{equation}
then there exists a constant $C=C(\alpha,\theta)$ such that 
$e^{t{\overline L}_\alpha} e^{\theta G}(x) \le e^{Ct}e^{\theta G}(x)$.
\end{prop}

\proof We note first that $\tL_\alpha$, defined in Eq. \eqref{lba}, 
for all $\alpha \in {\bf R}$, is the 
generator of a Markov process which we denote as $\tx(t)$. 
Indeed we have that
\bd
\tL_\alpha G(x) 
\,=\, \trace(\gamma T) - (1+2\alpha)r^2 \, \le \, C_1+ C_2 G(x)  
\ed
Since $G$ grows at infinity, $G$ is a Liapunov function for $\tx(t)$
and a standard argument \cite{Ka} shows that the Markov process
$\tx(t)$ is non-explosive.
Furthermore we have the bound 
\begin{eqnarray}
&&{\overline L}_\alpha \exp{\theta G}(x) \,=\, \nonumber \\
&&\,=\,\exp{\theta G(x)} \gamma
\left[ \trace( \theta T + \alpha I) +  
r( \theta^2 T - (1- 2\alpha)\theta  - \alpha (1-\alpha)T^{-1} ) r 
\right]  \nonumber \\
&&\,\le\, C \exp{\theta G(x)} \,,  \label{as}
\end{eqnarray}
provided $\alpha$ and $T_i$, $i=1,n$ satisfy the inequality 
\bd
\theta^2 T_i -(1-2\alpha)\theta -  \alpha (1-\alpha) T_i^{-1} \le 0 \,, 
\ed
or
\bd
-\alpha < \theta T_i < 1-\alpha \,.
\ed
We denote $\sigma_R$ as the exit time from the set $\{G(x) < R\}$,
i.e., $\sigma_R = \inf\{ t \ge 0, G(\tx(t)) \ge R \}$. If the initial
condition $x$ satisfies $G(x)=E<R$, we denote by $\tx_R(t)$ the process which
is stopped when it exits $\{G(x) < R\}$, i.e., $\tx_R(t) =\tx(t)$ for 
$t < \sigma_R$ and $\tx_R(t) = x(\sigma_R)$ for $t\ge \sigma_R$. 
Finally we set $\sigma_R(t) = \min \{\sigma_R, t\}$. 

By Eq. \eqref{as}, the function $W(t,x) = e^{-Ct}e^{\theta G(x)}$
satisfies the inequality $(\partial_t+{\overline L}_\alpha) W(t,x)\le
0$ and applying Ito's formula with stopping time to the
function $W(t,x)$ we obtain
\bd
\bE_x \left[ 
e^{-\int_0^{\sigma_R(t)} \left( (\alpha-\alpha^2)\gamma \tr T^{-1} \tr 
- \alpha \trace(\gamma I) \right)\, ds}  
e^{\theta G(\tx (\sigma_R(t)))} e^{-C \sigma_R(t) } \right] 
- e^{\theta G(x)} \le 0 \,,
\ed
and thus  
\bd
\bE_x \left[ 
e^{-\int_0^{\sigma_R(t)} \left( (\alpha-\alpha^2)\gamma \tr T^{-1} 
\tr - \alpha \trace(\gamma I) \right)\, ds } e^{\theta G(\tx (\sigma_R(t)))} 
\right] \,\le \,e^{C t} e^{\theta G(x)} \,.
\ed
Since the Markov process $\tx(t)$ is non-explosive $G(\tx_R(t)) 
\rightarrow G(\tx(t))$ almost surely as $R \rightarrow \infty$, 
so by the Fatou lemma we have 
\bd
e^{t{\overline L}_\alpha} e^{\theta G}(x) \,\le\, e^{Ct} e^{\theta G}(x)\,.
\ed
This concludes the proof of Lemma \ref{welldef}. \qed

The next three theorems are all consequences of the fact that
$\tL_\alpha$ is the generator of a Markov process which is similar to
the process generated by $L$: Indeed $L$ and $\tL_\alpha$ differ only
by the coefficient in front of the $r\nabla_r$ term. Therefore
repeating the proofs of \cite{RT2} we obtain

\begin{theorem} \label{feller} The semigroup $e^{t {\overline L}_\alpha}$ 
has a smooth kernel $q_\alpha(t,x,y)$ which belongs to 
${\mathcal C}^\infty((0,\infty)\times X \times X)$.
\end{theorem}

\proof The operator ${\tilde L}_\alpha$ satisfies the 
same H\"ormander-type condition that $L$ proven in \cite{RT2}, Proposition 
4.1.The result follows then from \cite{Ho} or \cite{No}. \qed

\begin{theorem} \label{irreducible} The semigroup $e^{t {\overline L}_\alpha}$
is positivity improving for all $t>0$.
\end{theorem}

\proof The semigroup $e^{t {\tilde L}_\alpha}$ is shown to be irreducible 
exactly as $e^{t L}$, see \cite{EPR2,RT2} using explicit computation
and the Support Theorem of \cite{SV}. The statement follows then from
the Feynman-Kac formula. \qed

As is apparent from the form of ${\overline L}_\alpha$ we will need
estimates on the observable $r^2$ in the sequel. Such estimates were
also crucial in \cite{RT2} for the construction of a Liapunov
function.

\begin{theorem} \label{tracking} 
Let $0\le \alpha < 1/k_2$ and let $t_E=
E^{1/k_2-1/2}$. There exists a set  of paths
\bd
S(x,E,t_E) \subset \{f \in \calC([0,t_E], X)\,;\, f(0)=x, G(x)=E\}\,,
\ed 
and constants $E_0 < \infty$ and $A,B,C>0$ such that for $E>E_0$
\bd
\bP \left\{ \tx  \in S(x,E,t_E) \right\} 
\,\ge\, 1 - A e^{ -B E^{2\alpha +1/2 -1/k_2}} \,,
\ed
and 
\begin{equation}\label{enes}
\int_0^{t_E}  \tr^2(s) \, ds \, \ge \, C E^{3/k_2 - 1/2}\,, 
{\quad} {\rm ~if~} \quad  \tx  \in S(x,E,t_E) \,.
\end{equation}
\end{theorem}

\proof The proof is exactly as in \cite{RT2}. 
One first sets $T_1=T_n=0$ in the equations of motion and then, by a
scaling argument, Theorem 3.3 of \cite{RT2}, one shows that the
deterministic trajectory satisfies the estimate \eqref{enes}. Then one
shows, see Proposition 3.7 and Corollary 3.8 of \cite{RT2}, that the
overwhelming majority of the random trajectories follows very
closely the deterministic ones. We refer the reader to \cite{RT2} for
further details. \qed

\begin{remark}\label{rem39}{\rm 
For large energy $E$, paths satisfying the bound \eqref{enes} have a
very high probability. From Eq. \eqref{enes} we obtain that, on a time
interval of order $1$,
\bd
\int_0^t  \tr^2(s) \ge C E^{2/k_2}\,,
\ed
for an overwhelming majority of the paths. 
}
\end{remark}

\begin{theorem} \label{liap} Let $t>0$ be fixed and suppose that 
$\alpha$ and $\theta$ satisfy the condition Eq.\eqref{condat}.  
There exist a constant $E_0$ and functions $\kappa(E)$ and $b(E)$ such that 
for $E> E_0$ 
\begin{equation}\label{kaa}
e^{t{\overline L}_\alpha} e^{\theta G}(x) \, \le \, \kappa(E) e^{\theta G(x)} 
+ b(E) {\bf 1}_{\{G\le E\}}(x)\,.
\end{equation}
Moreover there exist constants $c_1$ and $c_2$ such that  
\bd
\kappa(E)\le c_1 e^{-c_2 E^{2/k_2}}\,.
\ed 
\end{theorem}

\proof 
By Proposition \ref{welldef} the function $e^{t {\overline L}_\alpha}
e^{\theta G}(x)$ is bounded on any compact set. 
Therefore to show \eqref{kaa} it suffices to show that 
\bd
\sup_{\{x\,:\, G(x)>E\}} \bE_x \left[ 
e^{ - \int_0^t \left( \alpha (1-\alpha)\gamma \tr T^{-1} \tr 
-\alpha \trace(\gamma I) \right)\,ds} 
e^{ \theta \left( G(\tx(t)) - G(\tx) \right)} \right] \,\le \, \kappa(E) \,.
\ed 
Using Ito's formula we have 
\bd
G(\tx(t))-G(x)\,=\,\int_0^t \gamma (\trace(T)-\tr^2)\, ds  +
\int_0^t \sqrt{2\gamma T} 
\tr d\omega(s)\,,
\ed 
and thus we obtain
\begin{eqnarray}
&&\bE_x \left[ 
e^{ - \int_0^t \left( \alpha (1-\alpha) \gamma \tr T^{-1}\tr 
- \alpha \trace(\gamma I) \right)\,ds} e^{ \theta(G(x(t)) - G(x))} 
\right] \nn \\
&& = e^{t \gamma \trace(\theta T + \alpha I )} 
\bE_x \left[ e^{  - \int_0^t \tr \left( \alpha (1-\alpha) \gamma T^{-1}  - 
\gamma \theta  (1-2\alpha) \right) \tr \,ds} \times \right. \nn \\
&& \hspace{6cm} \left. \times  
e^{ \int_0^t \theta \sqrt{2\gamma T} \tr \, d\omega } \right]  \,.
\label{ex01}
\end{eqnarray}
Using the H\"older's inequality we find that the expectation on the
r.h.s of Eq. \eqref{ex01} can be estimated by 
\begin{eqnarray}
&& \bE_x \left[ e^{ -q \int_0^t \tr \left( \alpha (1-\alpha) \gamma T^{-1}  
- \gamma \theta  (1-2\alpha) \right) \tr \,ds }   
e^{ \frac{q p \theta^2 }{2} \int_0^t 
(\sqrt{2\gamma T}\tr)^2 \, ds }  \right]^{1/q} \nn \\
&& \quad \times  
\bE_x \left[e^{  -\frac{p^2\theta^2}{2} \int_0^t (\sqrt{2\gamma T}\tr)^2\, ds }
e^{  p \int_0^t \theta (2\gamma T)^{1/2} \tr \, d\omega )} 
\right]^{1/p} 
\nn \\
&& \,=\,   
\bE_x \left[ e^{ -q \gamma \int_0^t \tr \left( \alpha (1-\alpha)  T^{-1}  
- \theta (1-2\alpha)  +  p \theta^2 T \right) \tr\, ds }  \right]^{1/q} 
\,. \label{poo}
\end{eqnarray}
where we have used that the second factor is the 
expectation of a martingale with expectation $1$.

If $\theta$ and $\alpha$ satisfy the condition \eqref{condat}, then,
by choosing $p$ sufficiently close to $1$, the quadratic form in the
right side of Eq. \eqref{poo} is negative definite. Using  
Theorem \ref{tracking}  as in Theorem 3.11 of
\cite{RT2} we obtain
\begin{eqnarray}
&&\sup_{x \in U^C} \bE_x \left[ 
e^{  - \int_0^t \left( \alpha (1-\alpha) \tr T^{-1} \tr - 
\alpha \trace(\gamma I)\right)\,ds} e^{\theta( G(x(t)) - G(x))} \right] \nn \\
&& \,\le \, 
e^{ \gamma \trace(\theta T + \alpha I)}
e^{ -  C E^{2/k_2} \gamma \trace( \alpha (1-\alpha)  T^{-1}  -
(1-2\alpha)\theta +  p \theta^2 T) } \nn \\
&& \, \le \, c_1 e^{-c_2 E^{2/k_2}} \,. \nn
\end{eqnarray}
and this concludes the proof of Theorem \ref{liap}. \qed

As in Theorem \ref{compact} we obtain
\begin{theorem} \label{comp2} If $\alpha$ and $\theta$ satisfy the condition 
Eq.\eqref{condat}, then $e^{t{\overline L}_\alpha}$ extends to a
strongly continuous quasi-bounded semigroup on $\calH_{p,\theta}$ for
$1\le p< \infty$ and on $\calH^0_{\infty,\theta}$. Moreover 
$e^{t{\overline L}_\alpha}$ is compact on $\calH_{p,\theta}$, 
$1< p\le \infty$ and on $\calH^0_{\infty,\theta}$.
\end{theorem}

\proof The proof is a repetition of the proof of Theorem \ref{compact} and is 
left to the reader. \qed

As a consequence of Theorem \ref{comp2} and of the theory of semigroup of 
positive operators \cite{Gr} we obtain 

\begin{theorem} \label{ea1} 
If
\bd
\alpha \in \left( - \frac{ T_{\min} }{ T_{\max}  - T_{\min} } \,,\, 
1 + \frac{ T_{\min} }{ T_{\max}  - T_{\min} } \right)\,,
\ed
then 
\bd 
e(\alpha)\,=\, \lim_{t\to \infty} - \frac{1}{t} \log \Gamma^i_x(t,\alpha) 
\ed  
exists, is finite and independent both of $i$ and $x$. 
\end{theorem}

\proof By Theorem \ref{comp2}, $e^{t{\overline L}_\alpha}$ generates a 
strongly continuous semigroup on $\calH^{0}_{\infty,\theta}$ if
\begin{equation}\label{sd}
-\alpha < \theta T_i < 1-\alpha\,. 
\end{equation}
If $\alpha \le 0$, this implies that 
$|\alpha| <  \theta T_{\min} < \theta T_{\max} < 1+ |\alpha|$ and so
the set of $\theta$ we can choose is non-empty provided
\bd
\alpha > - \frac{ T_{\min} }{ T_{\max}  - T_{\min}}\,.
\ed
If $0<\alpha < 1$, we can always find $\theta$ such that \eqref{sd} is 
satisfied. Finally if $\alpha>1$ then \eqref{sd} implies that 
that 
\bd
\alpha < 1 + \frac{ T_{\min} }{ T_{\max}  - T_{\min}} \,.
\ed
By the definition of $R_i$, Eq. \eqref{ri}, $e^{-\alpha R_i} \in 
\calH^0_{\infty, \theta}$ since $-\alpha + \theta T_i <0$. 
Using now Lemma 2.7, we see that $\Gamma^i_x(t,\alpha)$ exists and is given by
\bd
\Gamma^i_x(t,\alpha) \,=\, e^{t(L-\alpha \sigma)}1(x)\,=\, 
e^{\alpha R_i} e^{t {\overline L}_\alpha}e^{- \alpha R_i}(x)\,.
\ed  
From Theorem \ref{irreducible} the semigroup $e^{t{\overline
L}_\alpha}$ is an {\em irreducible} semigroup of {\em compact}
operators on the Banach space $\calH^0_{\infty,\theta}$. From the
cyclicity properties of the spectrum of irreducible operators and from
the compactness it follows (see \cite{Gr}, Chapter C-III) that there
is exactly one eigenvalue $e^{-te(\alpha)}$ with maximal modulus and
this eigenvalue is real and simple. The corresponding eigenfunction
$f_\alpha$ is strictly positive and we denote as $P_\alpha$ the
one-dimensional projection on the eigenspace spanned by $f_\alpha$. In
particular if $g\ge 0$, then $P_\alpha g(x) >0$.

From compactness it follows that the complementary projection $(1-P_\alpha)$ 
satisfies the bound 
\begin{equation}\label{yu}
\left| e^{t{\overline L}_\alpha}(1-P_\alpha)f(x)\right| \, \le \, 
C e^{- t d(\alpha)}
\|f\|_{\infty,\theta} e^{\theta G(x)} \,. 
\end{equation}
for some constants $C>0$ and $d(\alpha) > e(\alpha)$ and for all $t>0$. 

From Lemma \ref{truc} and Eq. \eqref{yu} we obtain, for all $x \in X$, 
that  
\begin{eqnarray}
&&\lim_{t \to \infty} - \frac{1}{t} \log  \Gamma^i_x(t,\alpha) \nn \\
&& \,=\, 
\lim_{t \to \infty} - \frac{1}{t} \log  e^{t(L-\alpha \sigma_i)}1(x) 
\,=\, \lim_{t \to \infty} - \frac{1}{t} \log  e^{\alpha R_i} 
e^{t{\overline L}_\alpha} e^{-\alpha R_i}(x)  \\
&& \,=\, \lim_{t \to \infty} \left(- \frac{1}{t} \alpha R_i(x)\right) + 
e(\alpha) \nn\\
&&\quad + \lim_{t \to \infty} - \frac{1}{t} 
\log \left( P_\alpha e^{-\alpha R_i}(x) + 
 e^{te(\alpha)} 
e^{t{\overline L}_\alpha} ( 1- P_\alpha) e^{-\alpha R_i}(x)\right) \nn \\
&&\,=\, e(\alpha) \,. \nn
\end{eqnarray}
This concludes the proof of Theorem \ref{ea1}. \qed

Using now the identity \eqref{conj} we can prove the symmetry of
$e(\alpha)$. Theorem \ref{momgen} is then an immediate consequence of
the following result.

\begin{theorem}\label{ea2} 
If 
\begin{equation}\label{i99}
\alpha \in \left( - \frac{ T_{\min} }{ T_{\max}  - T_{\min} } \,,\, 
1+  \frac{ T_{\min} }{ T_{\max} - T_{\min}}  \right)\,,
\end{equation}
then
\bd
e(\alpha) = e(1-\alpha)\,.
\ed
\end{theorem}

\proof If $\alpha$ is in the interval \eqref{i99} and 
$-\alpha < \theta T_i < 1-\alpha$ 
then 
$e^{t{\overline L}_\alpha}$ is a strongly continuous compact semigroup on
$\calH^0_{\infty,\theta}$. By Lemma \ref{truc} 
\bd 
e^{t(L-\alpha\sigma_i)}\,=\, e^{\alpha R_i} e^{t{\overline L}_\alpha}
e^{-\alpha R_i}
\ed
is also a strongly continuous compact semigroup on the Banach space 
$\calH^0_{\infty,\theta,\alpha}=\{f\,;\, |f|e^{-\theta G + \alpha R_i} 
\in C_0(x)\}$ with the norm $\|f\|_{\infty,\theta,\alpha}=
\sup|f|e^{\theta G+\alpha R_i}$.  

The dual semigroup $(e^{t(L-\alpha\sigma_i)})^*$ is a compact
semigroup on the Banach space (of measures)
$(\calH^0_{\infty,\theta,\alpha})^*$. By Theorem
\ref{irreducible} $(e^{t(L-\alpha\sigma_i)})^*$ maps 
$(\calH^0_{\infty,\theta,\alpha})^*$ into measures with smooth
densities and on densities $(e^{t(L-\alpha\sigma_i)})^*$ acts as
\bd
(e^{t(L-\alpha\sigma_i)})^*(\rho(x)dx)
\,=\,(e^{t(L^T-\alpha\sigma_i)}\rho(x))dx\,.
\ed
By Lemma \ref{detbal} we have 
\begin{equation}\label{kl}
e^{-R_i} e^{t(L-(1-\alpha)\sigma_i)}1(x)\,=\, J e^{t(L^T-\alpha \sigma_i)}J
e^{-R}(x) \,.
\end{equation}
Since $-\alpha < \theta T_i < 1-\alpha$, $e^{-R_i}$ is a density of a
measure in $(\calH^0_{\infty,\theta,\alpha})^*$.  Since
$(e^{t(L-\alpha\sigma_i)})^*$ is compact and irreducible with spectral
radius $e(\alpha)$ we obtain using Eq. \eqref{kl}
\begin{eqnarray}
e(\alpha) \,&=&\, \lim_{t \to \infty} -\frac{1}{t} \log 
\| J (e^{t (L^T -\alpha \sigma_i)}J e^{-R_i})dx \| \nn \\
\,&=&\, \lim_{t \to \infty} -\frac{1}{t} \log \left( 
\sup_{ f \le e^{\theta G + \alpha R_i}} \int f e^{-R_i} 
e^{t(L-(1-\alpha)\sigma_i)}1\, dx \right) \,,\nn \\
\,&=&\, e(1-\alpha) \,.\nn
\end{eqnarray}
In the last equality we have used Theorem \ref{ea1} and the fact that
$f e^{-R_i}$ is a finite measure. This concludes the proof of Theorem
\ref{ea2}. \qed

We finally obtain the Gallavotti-Cohen fluctuation theorem

\begin{theorem}\label{gc} 
There is a neighborhood $O$ of the interval 
$[-\langle \sigma \rangle_{\nu}\,, \langle \sigma \rangle_{\nu}]$ such
that for $A \subset O$ the fluctuations of $\sigma_i$ in $A$ satisfy
the large deviation principle with a large deviation functional $I(w)$
obeying
\bd
I(w)-I(-w)\,=\, -w\,,
\ed
i.e., the odd part of $I$ is linear with slope ${-1/2}$.  
\end{theorem}

\proof First we note that $e(\alpha)$ is a real analytic function since it 
is identified with an eigenvalue of a compact operator. 
A simple computation gives that 
\bd
\left.\frac{d}{d\alpha}e(\alpha)\right|_{\alpha=0}\,=\, 
\langle \sigma \rangle_\nu \,.
\ed
The function $e(\alpha)$ is analytic and convex. By the result of
\cite{EPR2} it is not identically zero, and so the symmetry the
symmetry $e(\alpha)=e(1-\alpha)$ implies that the set of the values of
$\frac{d}{d\alpha}e(\alpha)$ is a neighborhood of $[-\langle \sigma
\rangle_{\nu}\,, \langle \sigma \rangle_{\nu}]$.

The large deviation principle is a direct application of the 
G\"artner-Ellis theorem, \cite{DZ}, Theorem 2.3.6. 
The large deviation functional is given by the Legendre transform of 
$e(\alpha)$ and so we have 
\begin{eqnarray}
I(w) \,&=&\, \sup_{\alpha} \left\{ e(\alpha) - \alpha w \right\}
\,=\, \sup_{\alpha}\left\{ e(1-\alpha) - \alpha w \right\}  \nn \\
\,&=&\, \sup_{\beta} \left\{ e(\beta) - (1- \beta) w \right\}  
\,=\, I(-w) - w \,. \nn
\end{eqnarray}
\qed


\begin{thebibliography}{99}

\bibitem{BM} Balaji, S. and Meyn, S. P.:
\newblock  Multiplicative ergodicity and large deviations for an irreducible 
Markov chain.  
\newblock Stoch. Proc. Appl. {\bf 90} 123--144 (2000).


\bibitem{Cr} Crooks, G.E.:
\newblock Path-ensemble averages in systems driven far from equilibrium.
\newblock Phys. Rev. E {\bf 61}, 2361--2366 (2000)


\bibitem{DS}Deuschel, J.-D. and Stroock, D.W.: 
\newblock {\em Large deviations}. 
\newblock Pure and Applied Mathematics {\bf 137}. 
Boston:  Academic Press, 1989


\bibitem{DZ}Dembo, A. and Zeitouni, O.: 
\newblock {\em Large deviations techniques and applications}.  
\newblock Applications of Mathematics {\bf 38}. New-York: Springer-Verlag 1998 


\bibitem{EH}
Eckmann, J.-P. and Hairer, M.:
\newblock Non-equilibrium statistical mechanics of strongly anharmonic 
chains of oscillators. 
\newblock  Commun. Math. Phys. {\bf 212}, 105--164 (2000)


\bibitem{EPR1}
Eckmann, J.-P., Pillet C.-A., and Rey-Bellet, L.:
\newblock Non-equilibrium statistical mechanics of anharmonic chains
coupled to two heat baths at different temperatures.
\newblock Commun. Math. Phys. {\bf 201}, 657--697 (1999)


\bibitem{EPR2}
Eckmann, J.-P., Pillet, C.-A., and Rey-Bellet, L.:
\newblock Entropy production in non-linear, thermally driven
Hamiltonian systems.
\newblock J. Stat. Phys. {\bf 95}, 305--331 (1999)


\bibitem{ECM} 
Evans, D.J., Cohen, E.G.D., and Morriss, G.P.:
\newblock Probability of second law violation in shearing steady flows. 
\newblock Phys. Rev. Lett. {\bf 71}, 2401--2404 (1993)


\bibitem{GC}
Gallavotti, G. and Cohen E.G.D.: 
\newblock Dynamical ensembles in stationary states. 
\newblock J. Stat. Phys. {\bf 80}, 931--970 (1995)


\bibitem{Ge} Gentile, G.:
\newblock Large deviation rule for Anosov flows.  
Forum Math. {\bf 10} 89--118 (1998)


\bibitem{Gr} Greiner,~G: 
\newblock Spectral theory  of positive semigroups on Banach lattices. 
\newblock In {\em One-parameter semigroups of positive operators} Lecture 
Notes in Mathematics {\bf 1184}, Ed. R. Nagel, Berlin: Springer, 1986, 
pp 292--332 


\bibitem{Ho}
H\"ormander, L.:
\newblock {\em The Analysis of linear partial differential operators}.
\newblock Vol {\bf III}, Berlin: Springer, 1985


\bibitem{JP1} Jaksic V. and Pillet C.-A.: 
\newblock On entropy production in quantum statistical mechanics. 
\newblock Commun. Math. Phys. {\bf 217}, 285--293 (2001)


\bibitem{JP2} Jaksic V.and Pillet C-A.: 
\newblock Non-equilibrium steady states of finite quantum systems coupled to 
thermal reservoirs.
\newblock Preprint (2001)


\bibitem{Ja} Jarzynski, C.: 
\newblock Hamiltonian derivation of a detailed fluctuation theorem. 
\newblock J. Statist. Phys.  {\bf 98}, 77--102 (2000)


\bibitem{Ka}
Has'minskii, R.Z.:
\newblock {\em Stochastic stability of differential equations}. 
\newblock Alphen aan den Rijn---Germantown: Sijthoff and Noordhoff, 1980


\bibitem{Ku} 
Kurchan, J:
\newblock Fluctuation theorem for stochastic dynamics. 
\newblock J. Phys.{\bf A 31}, 3719--3729 (1998)


\bibitem{Ma1} Maes, C. 
\newblock The fluctuation theorem as a Gibbs property. 
\newblock J. Stat. Phys. {\bf 95} 367--392 (1999)


\bibitem{Ma2} Maes, C. 
\newblock Statistical mechanics of entropy production: Gibbsian hypothesis 
and local fluctuations.
\newblock Preprint (2001)


\bibitem{Ma3} Maes, C., Redig, F., and Verschuere, M. 
\newblock No current without heat
\newblock Preprint (2000)


\bibitem{MeTw} 
Meyn, S.P. and Tweedie, R.L.:
\newblock {\em Markov Chains and Stochastic Stability.} 
\newblock Communication and Control Engineering Series, London: 
Springer-Verlag London, 1993 


\bibitem{LeSp}
Lebowitz, J.L. and Spohn, H.: 
\newblock A Gallavotti-Cohen-type symmetry in the large deviation 
functional for stochastic dynamics. 
\newblock J. Stat. Phys. {\bf 95}, 333-365 (1999)


\bibitem{No} Norriss, J.:
\newblock Simplified Malliavin Calculus.  
\newblock In {\em S\'eminaire de probabilit\'es XX}, 
Lectures Note in Math. {\bf 1204}, 0 Berlin: Springer, 1986, pp. 101--130


\bibitem{P} Pillet, C.-A.: 
Entropy production in classical and quantum systems. 
Markov Proc. Relat. Fields {\bf 7}, 145-157, (2001). 


\bibitem{RT1}
Rey-Bellet, L. and Thomas, L.E.:
\newblock Asymptotic behavior of thermal non-equilibrium steady states 
for a driven chain of anharmonic oscillators.
\newblock Commun. Math. Phys. {\bf 215}, 1--24 (2000) 


\bibitem{RT2}
Rey-Bellet, L. and Thomas, L.E.:
\newblock Exponential convergence to non-equilibrium stationary states in 
classical statistical mechanics 
\newblock To appear in Commun. Math. Phys.


\bibitem{Ru1} Ruelle, D.:
\newblock Entropy production in quantum spin systems.
\newblock Preprint (2000)
 





\bibitem{SV}
Stroock, D.W. and Varadhan, S.R.S.:
\newblock On the support of diffusion processes with applications to the
strong maximum principle.
\newblock In {\em Proc. 6-th Berkeley Symp. Math. Stat. Prob.}, Vol {\bf III}, 
Berkeley: Univ. California Press, 1972, pp. 361--368

\bibitem{Wu}
Wu,~L.:
\newblock Uniformly integrable operators and large deviations for Markov 
processes.
\newblock J. Funct. Anal. {\bf 172}, 301--376 (2000)

\end{thebibliography}
\end{document}